# Machine learning method for $^{12}$C event classification and reconstruction in the active target time-projection chamber

Huangkai Wu[a,b], Youjing Wang[c], Yumiao Wang[c], Xiangai Deng[c,d], Xiguang Cao[a,b,e], Deqing Fang[c], Weihu Ma[c], Hongwei Wang[a,b,e], Wanbing He[c,*], Changbo Fu[c,*], Yugang Ma[c,d,*]

[a]*Shanghai Institute of Applied Physics, Chinese Academy of Sciences, 201800, Shanghai, China*
[b]*University of Chinese Academy of Sciences, 100049, Beijing, China*
[c]*Key Laboratory of Nuclear Physics and Ion-beam Application (MOE), Institute of Modern Physics, Fudan University, 200433, Shanghai, China*
[d]*Shanghai Research Center for Theoretical Nuclear Physics, NSFC and Fudan University, 200438, Shanghai, China*
[e]*Shanghai Advanced Research Institute, Chinese Academy of Sciences, 201210, Shanghai, China*

**Abstract**

Active target time projection chambers are important tools in low energy radioactive ion beams or gamma rays related researches. In this work, we present the application of machine learning methods to the analysis of data obtained from an active target time projection chamber. Specifically, we investigate the effectiveness of Visual Geometry Group (VGG) and the Residual neural Network (ResNet) models for event classification and reconstruction in decays from the excited $2_2^+$ state in $^{12}$C Hoyle rotation band. The results show that machine learning methods are effective in identifying $^{12}$C events from the background noise, with ResNet-34 achieving an impressive precision of 0.99 on simulation data, and the best performing event reconstruction model ResNet-18 providing an energy resolution of $\sigma_E < 77$ keV and an angular reconstruction deviation of $\sigma_\theta < 0.1$ rad. The promising results suggest that the ResNet model trained on Monte Carlo samples could be used for future classifying and predicting experimental data in active target time projection chambers related experiments.

## 1. Introduction

Radioactive ion beams (RIBs) and gamma beams play important roles in modern nuclear physics studies [1–4]. However, the beams have their own shortages. Such as, beam intensities are too weak, or lifetimes of RIBs are too short etc. To overcome them, the technology of active target time projection chamber (AT-TPC) has been developed in the past two decades [5, 6], which has revolutionized the fields. Typically, AT-TPCs utilize various gases serving as targets and detectors at the same time, which enables them to offer high-resolution and efficient reconstruction of charged particle trajectories and energies. This feature enables AT-TPCs as a popular choice in nuclear physics research, particularly in experiments involving RIBs and gamma beams. Several AT-TPCs have been developed in recent years, such as the MAIKo by Kyoto University for investigating shell evolution and cluster structure [7], TexAT by Texas A&M University for the study of shell evolution [8], and MATE by the Institute of Modern Physics, Chinese Academy of Sciences for investigating heavy-ion fusion reactions at stellar energies [9]. For more information on AT-TPCs, one can refer to a review article by Bazin *et al.* [5, 6].

Although AT-TPCs offer many advantages, analyzing the large amount of data they generate presents a significant challenge. For instance, a typical week-long experiment produces approximately 10 terabytes of raw data, which must be processed to obtain charge deposition and spatial information. Conventional analysis methods typically rely on manual inspection and event selection, which is time-consuming and subject to human bias. However, convolutional neural networks (CNNs) have demonstrated great potential in automating the analysis of particle detector outputs, including event reconstruction and particle identification, and could offer a solution to this challenge.

Machine learning methods have been extensively applied in different fields in recent years [10–15]. Some reviews are available for nuclear physics [16–18]. It was also applied in some particle detectors and facilities successfully, for example [19–22], including the AT-TPC [23], TexAT [24] JUNO [25], and NeuLAND [26] etc., to identify particle tracks and reconstruct energy and vertex information. For examples, researchers have used a pre-trained network, VGG-16, trained on ImageNet, to classify proton tracks on simulated and experimental datasets with impressive results [23]. JUNO [27–29] has applied several machine learning approaches, such as Boosted Decision Trees (BDT), Deep Neural Networks (DNN), VGG-17, and ResNet-53, to vertex and energy reconstruction on simulated datasets [25, 30]. DNNs have also successfully reconstructed total energy and position directly from raw digitized waveforms on the

EXO-200 experiment [31].

In this article, we explore the application of using CNNs to analyze the simulation data from AT-TPCs. The motivation behind this analysis is to investigate the properties of $^{12}$C [32–35] particularly its Hoyle state. The Hoyle state, which was first raised by Fred Hoyle in 1953 [32], has been a subject of interest in nuclear physics due to its unique structure and its significant role in the nucleosynthesis in stars [36–40].

After the discovery of the Hoyle state, Morinaga *et al.* [41] proposed that studying the rotational band built on top of it could provide valuable insights into the structure of the Hoyle state. This suggestion sparked continuous research interest in the Hoyle state rotational band [42–46]. Recently, Zimmerman *et al.* identified the first excited state of the Hoyle state band unambiguously by using the O-TPC [47], a cutting-edge active target detector at the *HIγS* facility [48]. However, the nature of the $2_2^+$ resonance state, which is located approximately 2 MeV above the Hoyle state in $^{12}$C, is not yet fully understood. Additional research is highly needed to fully characterize the strength of $2^+$ at higher excitation energies, as well as the relationship between the $0_3^+$ state and the $2^+$ state [49].

This article will be organized as following: Section 2 will present the procedure for preparing and preprocessing the training data to be used in CNN models for event classification and reconstruction. In the Section 3.1, we discuss the performance of different CNN models in event classification, while a detailed discussion of event reconstruction can be found in the Section 3.2. Finally, in Section 4, we present the conclusions of this work and possible directions for future researches.

## 2. Dataset

The present numerical studies based on the fMata-TPC (Fudan Multi-purpose Active TArget Time Projection Chamber) currently under construction at Fudan University, Shanghai, China. The TPC has a sensitive area of $144 \times 288$ mm$^2$, consisting of $32 \times 64$ rectangle pixels that increase in size from the inner to the outer region, as shown in Figure 1. The pixel size in the *x*-direction (short side) increases from 2 mm to 6 mm, while it remains as a constant of 4.5 mm in the *z*-direction (long side).

### 2.1. Data preparation

The training and testing of CNN are based on Monte Carlo samples generated by the Actarsim[51] . The Actarsim is built upon the Geant4[52] and Garfield++[53] frameworks for simulating the Active Target Time-Projection Chamber (AT-TPC).

In this work, we will focus on the feasibility of studying the $^{12}$C Hoyle rotation band by using an AT-TPC at the SLEGS facility [54, 55] The simulation employed beams of gamma rays with energies range of 9.2 to 10.8 MeV to induce photonuclear reactions with 30 mbar iso-C$_4$H$_{10}$ gas in the chamber. As shown in the Fig. 2, three different decay channels are included in the $^{12}$C dataset: $^{12}$C$(\gamma, \alpha_0)^8$Be$_{g.s.}$ sequential decay, $^{12}$C$(\gamma, \alpha_1)^8$Be$_{2^+}$ sequential decay, and $^{12}$C$(\gamma, 3\alpha)$ direct decay. The dissociation of $^{12}$C into an $\alpha$ particle and $^8$Be$_{g.s.}$ accounts for 98% of the dataset [43].

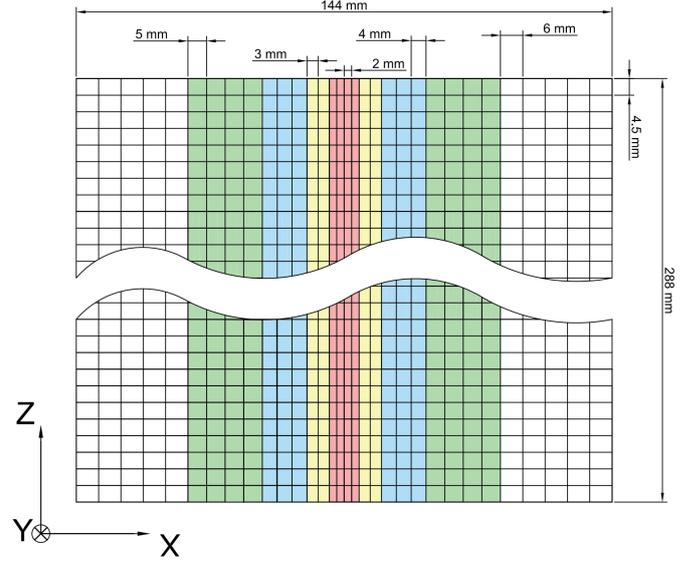

Fig. 1. (Color online) The readout board, a micromagas plane[50], of fMata-TPC with a sensitive area of $144 \times 288$ mm$^2$. The sensitive area is splited into $32 \times 64$ rectangle pixels with different sizes. The pixel sizes in the *x*-direction (short side) are 2 mm, 3 mm, 4 mm, 5 mm, and 6 mm, while it keeps as a constant of 4.5 mm in the *z*-direction (long side).

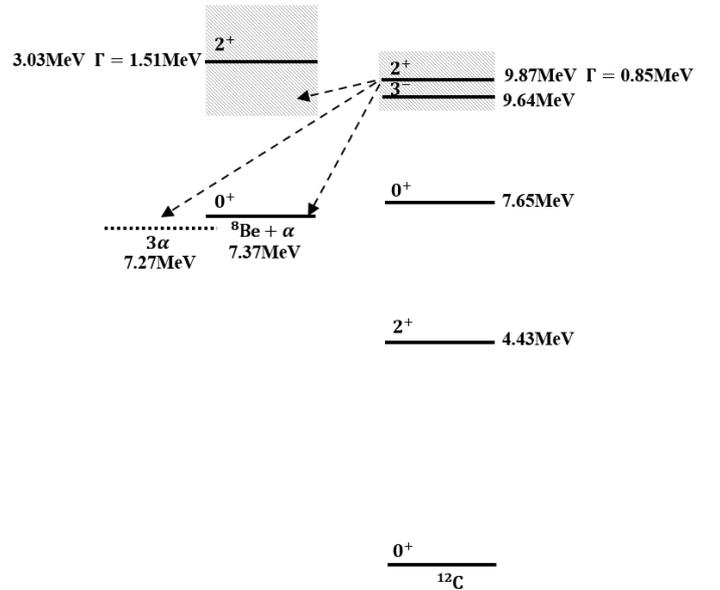

Fig. 2. Decay Scheme of the $2_2^+$ state in $^{12}$C's Hoyle state band. The $2_2^+$ state may decay to the ground state, $2^+$ excited state of $^8$Be, or triple $\alpha$ directly.

However, despite efforts to maintain air-tightness and remove the gas adsorbed on the chamber's inner surface, some air may still leak into the chamber from the outside, resulting in undesired photonuclear reactions that were recorded as background. As a consequence, the simulation produced four sets of data ($^{12}$C, $^{13}$C, $^{14}$N, and $^{16}$O), which collectively represent potential experimental background and decay events of interest. Each dataset records the charge deposited and electron drift time of each pixel event. A summary of the reaction channels considered in this study is shown in Table 1. It should be noted that, in principle, the influence of the 3⁻ state of $^{12}$C should be taken into account. However, due to the small E3 transition probability, this level was not included in the simulation. Additionally, the reaction channel of $^{13}$C $(\gamma, n)^{12}$C was not considered, as the resulting $^{12}$C trace in the TPC would be short and easily distinguishable from other events. On the other hand, all levels that could be populated by gamma beams with energies ranging from 9.2 to 10.8 MeV for $^{14}$N and $^{16}$O nuclei were included in the simulation.

| Reaction | $J^\pi$ | $^{12}$C level(MeV) | Width(MeV) |
|---|---|---|---|
| $^{12}$C$(\gamma, \alpha)^8$Be$_{g.s.}$ | $2_2^+$ | 9.870 | 0.850 |
| $^{12}$C$(\gamma, \alpha)^8$Be$^*(2^+, 3.03 MeV)$ | $2_2^+$ | 9.870 | 0.850 |
| $^{12}$C$(\gamma, 2\alpha)\alpha$ | $2_2^+$ | 9.870 | 0.850 |
| $^{13}$C$(\gamma, \alpha)^9$Be | $1/2^+$ | 10.996 | 0.037 |
| $^{14}$N$(\gamma, p)^{13}$C | $1^+$ | 9.703 | 0.015 |
|  | $2^+, 1^+$ | 10.101 | 0.012 |
|  | $1^{(-)}$ | 10.226 | 0.080 |
|  | $2^+$ | 10.432 | 0.033 |
| $^{16}$O$(\gamma, \alpha)^{12}$C | $1^-$ | 9.585 | 0.420 |
|  | $2^+$ | 9.844 | 0.620 |

Table 1: The summary of nuclear reactions which are considered in the MC simulation in this work.

## 2.2. Dataset split

In this study, each dataset contained $10^5$ events, uniformly distributed within the active area of the TPC. 75% of each dataset was reserved for iterative optimization of the model parameters, while another 5% was used to validate the performance of the models after each epoch. The remaining 20% was used to estimate the performance after the end of training. The primary objective of this study was to distinguish $^{12}$C events from the background and reconstruct the energy, emission angle, and vertex of the $^{12}$C events.

## 2.3. Data pre-processing

One of the notable features of TPC is its ability to capture the three-dimensional trajectory of charged particles, where the x and z dimensions are obtained from the pixels on the readout board, while the third dimension is derived from the electron drift time. To process the dataset using planar CNNs, the study adopted a two-dimensional projection method. This involved projecting the 3D tracks onto the readout board plane and dividing the plane into 2048 pixels. The data was then be represented as 32x64 pixel images with two channels, one for the deposited charge and another one for the electron drift time. This representation is similar to the RGB color model used in images. While it is possible to train the model on point-cloud data using 3D CNNs, this approach requires more computational resources.

## 3. Planar CNN models

CNN is a sophisticated machine learning model that are designed to identify and extract features from complex, spatially dependent data, like those from particle detectors. They employ a series of convolutional layers to extract features from the input data, which are then fed into fully connected layers to perform classification or regression tasks. Compared to traditional machine learning algorithms, CNNs have numerous benefits.

One of the most significant advantages of CNNs is their ability to extract localized features. This means that they are able to identify features at different scales and orientations, making them well-suited for analyzing particle detector data, which may contain intricate spatial patterns.

Furthermore, CNNs have an end-to-end learning capability, allowing them to learn to perform the entire analysis task from raw data to the final output. This feature eliminates the need for manual feature engineering or pre-processing, streamlining the analysis process and making it more automated.

Additionally, CNNs are highly robust to noise, making them suitable for analyzing data with a high degree of uncertainty or noise. The ability to handle noisy data can be essential in many scientific and industrial applications, including particle physics, medical diagnosis, and image recognition.

### 3.1. Classification

In this work, two well-known CNN architectures, the Visual Geometry Group (VGG)[56] and the Residual neural Network (ResNet)[57], are employed for our purpose. VGG is a classic CNN architecture that has been widely used for image recognition tasks, while ResNet is a more recent architecture that has achieved state-of-the-art performance on various computer vision tasks. The performance of both architectures will been compared on the specific task.

#### 3.1.1. $^{12}$C events identification

Event and background classification were performed using VGG-16, ResNet-18, and ResNet-34 models. To prevent over-fitting, regularization methods were applied during training, and the relevant hyperparameters for each model were listed in Table 2. To monitor the effect of over-fitting, a learning curve was drawn, as shown in Fig. 3. This curve tracks the value of the model's loss and accuracy on the training and validation sets over time. To evaluate the models' performance, the precision, recall, and F1-score metrics were used. Compared to classification accuracy, these metrics provide more nuanced insights into a model's performance and error characteristics. The focus of this section was on identifying $^{12}$C decay events from background, and these metrics allowed for a precise assessment of the models' ability to achieve so.

The multi-class classification results from the ResNet-34 model are presented using a confusion matrix in Fig. 4. The figure aids in identifying which background parts have a significant impact on $^{12}$C event classification. Table 3 summarizes the results of each model after testing, indicating that ResNet-34 achieved an impressive F1 score of 0.9886 with fewer training epochs. Its precision is 99.51%, suggesting a low probability of error in identifying $^{12}$C events. Its recall is 98.23%, suggesting that only 1.77% of $^{12}$C decay events were not successfully classified by the ResNet-34 model. This study holds great potential

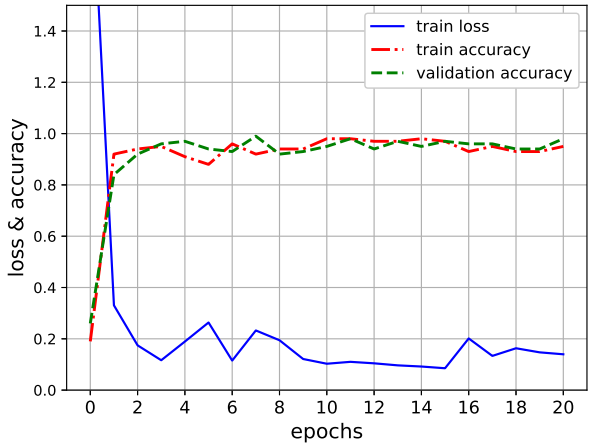

**Fig. 3.** (Color online) A learning curve depicting the value of the model's loss and accuracy on samples in the training set and the validation set over time. The *x*-axis indicates epochs of training, where one epoch represents one complete pass through the dataset for stochastic gradient descent.

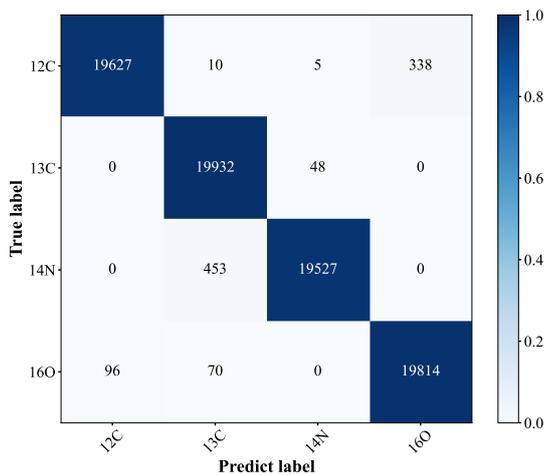

**Fig. 4.** (Color online) Confusion matrix for $^{12}$C events and background classification result from ResNet-34 model. Each column represents the class predicted by the classifier, while each row represents the true class label. Detailed results are shown in Table 3

for the successful application of the model to experimental data classification after training. This means that experimenters no longer need to manually categorize experimental data event-by-event.

It should be acknowledged that the results reported in Table 3 represent the upper limits that can be achieved by these three plane models. This is because the training and testing datasets used do not contain the structural noise that is typically present in real experimental data.

### 3.1.2. Classification of decay channels

The study of Hoyle state configuration has long been focused in nuclear physics, and in recent years, there has been a growing trend of studying the decay width of the Hoyle state's direct $3\alpha$ decay channel to gain insights into its configuration[42]. Among the decay modes of the excited $2_2^+$ state of Hoyle state band, direct $3\alpha$ decay is a rare decay mode, and there is currently no definitive experimental evidence for its observation. In this decay mode, the three outgoing $\alpha$ particles have no correlation information, have equal energies in the center-of-mass frame, and form 120-degree angles with each other, as illustrated in Fig. 5. To address this issue, we aim to utilize machine learning to classify the decay type of the $2_2^+$ state, and then provide guidance for measuring this decay channel in experiments.

| Parameters | Values |
|---|---|
| Batch size | 100 |
| Learning rate | 0.0001 |
| Loss | CrossEntropyLoss |
| Optimizer | Adam ($\beta_1 = 0.9, \beta_2 = 0.999$) |
| Weight decay | $\begin{cases} 0, & \text{for VGG}-16 \\ 10^{-4}, & \text{for ResNet}-18 \\ 10^{-4}, & \text{for ResNet}-34 \end{cases}$ |
| Scheduler type | ReduceLROnPlateau (mode='min',factor=0.1,patience=2) |

Table 2: Inputing hyperparameters for VGG-16, ResNet-18 and ResNet-34.

| Algorithm | Epochs | Precision | Recall | F1 |
|---|---|---|---|---|
| VGG-16 | 20 | 0.9523 | 0.8633 | 0.9056 |
| ResNet-18 | 10 | 0.9324 | 0.9195 | 0.9259 |
|  | 20 | 0.9642 | 0.9488 | 0.9565 |
| ResNet-34 | 10 | 0.9951 | 0.9823 | 0.9886 |

Table 3: CNN model results: simulated data from Geant4 was used for training and testing each of CNN models.

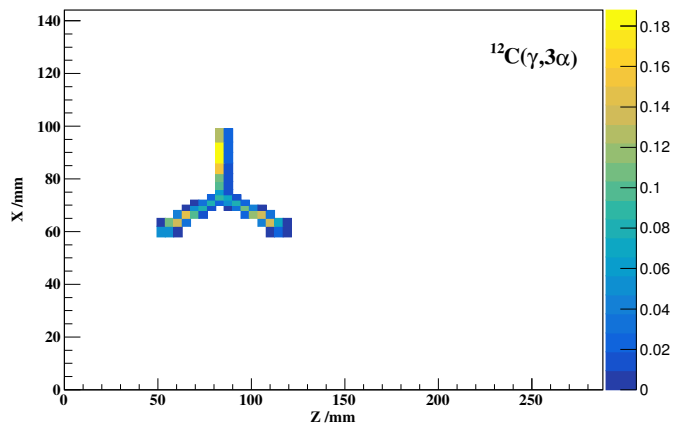

**Fig. 5.** (Color online) A typical trace of direct $3\alpha$ decay events simulated by Actarsim for $E_\gamma = 10.03$ MeV on fMata-TPC. The angles between two of the triple $\alpha$ are 120° in their center-of-mass coordinate. The deposited energy (in MeV) in each pixel is presented by a color scale.

As shown in Fig. 6, the direct decay can be well effectively distinguished from the other two decay types, especially when compared with the sequential decay channel $^{12}$C$(\gamma, \alpha_0)^8$Be$_{g.s.}$. However, the decay of $^{12}$C$(\gamma, \alpha_1)^8$Be$_{2^+}$ can slightly confuse the classification of direct $3\alpha$ decay. It should be noticed that a stringent upper limit on the direct decay branching ratio of $\frac{\Gamma_{3\alpha}}{\Gamma} < 5.7 \times 10^{-6}$ has been obtained in the most recent study[43]. Therefore, minor confusion has a huge impact on the measurement of $3\alpha$ decay events.

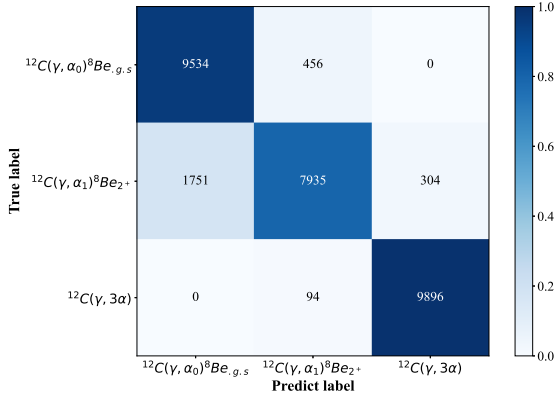

**Fig. 6.** (Color online) Confusion matrix for the decay modes of the possible excited state on Hoyle state band with ResNet18 approach. Each column is the decay channel predicted by the classifier, and each row is the true decay channel of the event.

### 3.2. Regression

In the following section, we evaluate the performance of the VGG-16, ResNet-18 and ResNet-34 models for reconstructing the primary vertex, emission angle and energy of decay products. The the reconstruction performance is analyzed as a function of the difference between the predicted and true values. Two characteristics, bias and $\sigma$, which are defined by Gaussian fit, are used to measure the performance of the trained models on the testing dataset.

#### 3.2.1. Vertex reconstruction

As shown in Fig. 7, ResNet-18 has a reconstruction bias of less than 0.4 mm on the three dimensions of the primary vertex (x, y, z), and the reconstruction resolution is within an acceptable range. Specifically, the $\sigma_z$ value of 1.12 mm is roughly equal to the spatial resolution in the z direction (approximated as $4.5/\sqrt{12}$ mm). In the x direction, the pixel size ranges from 2 mm to 6 mm, as shown in Fig. 1. Consequently, $\sigma_x$ falls between $2/\sqrt{12}$ and $6/\sqrt{12}$. In the TPC, the position y is determined by the product of electron drift velocity and drift time. As a result, $\sigma_y$ can be roughly equal to the deviation of electron longitudinal diffusion, which can be estimated using $\sigma_{diff} = \sqrt{2D_L h/v_{drift}}$, where $D_L$ is the longitudinal diffusion coefficient, $h$ is the vertical height, and $v_{drift}$ is the electron drift velocity. The value of $\sigma_{diff}$ typically ranges from 1-2mm.

Tables (5) and (4) summarize the performance of CNNs in primary vertex reconstruction. The results presented in Table 5 indicate that VGG-16 underperforms due to suboptimal data processing, whereas ResNet-18 and ResNet-34 demonstrate comparable and strong capabilities for vertex reconstruction.

| Parameters | Values |
|---|---|
| Batch size | 32 |
| Learning rate | 0.0001 |
| Loss | MSELoss |
| Optimizer | Adam ($\beta_1 = 0.9, \beta_2 = 0.999$) |
| Weight decay | $\begin{cases} 0, & \text{for VGG}-16 \\ 0.1, & \text{for ResNet}-18 \\ 0.15, & \text{for ResNet}-34 \end{cases}$ |
| Scheduler type | ExponentialLR (optimizer,gamma=0.35) |

Table 4: Hyperparameters for models on vertex reconstruction.

| Algorithm | epochs | $\sigma_x$ (mm) | $\sigma_y$ (mm) | $\sigma_z$ (mm) |
|---|---|---|---|---|
| VGG-16 | 30 | 2.21 | 2.48 | 5.54 |
| ResNet-18 | 15 | 0.71 | 0.81 | 1.12 |
| ResNet-34 | 15 | 0.78 | 0.78 | 1.76 |

Table 5: The standard deviation of vertex reconstruction for models on the $^{12}$C test dataset.

However, ResNet-18 consumes fewer resources than ResNet-34. Thus, ResNet-18 offers the best balance between performance and efficiency for this task.

#### 3.2.2. Energy reconstruction

To achieve a thorough reconstruction of the $^{12}$C decay, it is crucial to precisely reconstruct the energy and emission angle of the decay products. The use of CNNs has proven effective in this task. However, accurate reconstruction of energy and momentum requires preprocessing the dataset. This is mainly due to the experimental difficulty in distinguishing between the three alpha particles that decay from $^{12}$C, making it experimentally challenging to differentiate between them. To improve learning accuracy and reduce the number of model parameters being trained, a smart approach is employed: sorting the three $\alpha$ particles in descending order based on their energies.

Fig. 8 illustrates the distribution of the discrepancy between the predicted and true energies to evaluate the performance of energy reconstruction. The bias of these distributions is quite small, at less than 10 keV. To compare the performance of VGG and ResNet in this task, we summarized the results in Table 7.

The table indicates that ResNet outperforms VGG-16, achieving a reconstructed energy resolution of less than 80 keV, which is a remarkable accomplishment in the field of photodisintegration product energy reconstruction. Finally, Table 6 presents the hyperparameters used for training the CNN models in energy reconstruction.

#### 3.2.3. Trace angular reconstruction

The CNNs also were applied to reconstruct the emission angle of decay products. However, due to the decay channels of $^{12}$C($\gamma, \alpha_0$)$^8$Be$_{.g.s}$ and $^{12}$C($\gamma, \alpha_1$)$^8$Be$_{2+}$, which produce the two $\alpha$ particles that decay from $^8$Be, reconstructing emission angles is particularly challenging because of the relatively lower energies of these particles and their shorter tracks in the TPC. As a result of these challenges, the CNN model's performance was poor, necessitating additional data processing steps before angle reconstruction could be carried out effectively.

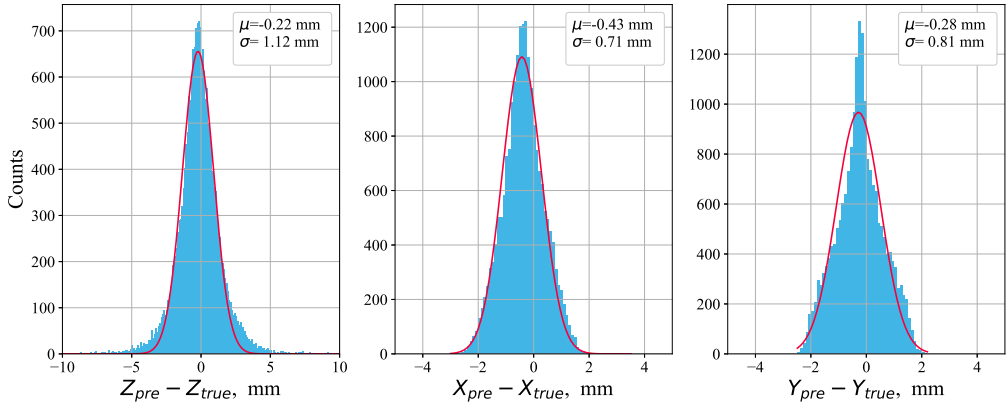

**Fig. 7.** The distribution of the reaction vertex residues ($\mathbf{r}_{prediction} - \mathbf{r}_{true}$) in $x, y, z$ dimensions. The predictions are produced by ResNet-18.

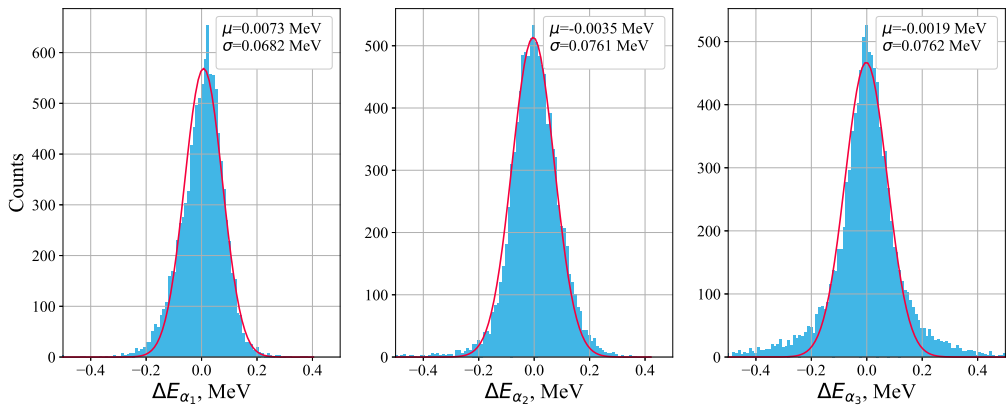

**Fig. 8.** (Color online) The distribution of the energy residues ($E_{prediction} - E_{true}$), where $E_{prediction}$ is reconstructed energy by using $^{12}$C test dataset and the ResNet-18 architecture. The left panel is the result for $\alpha$ particles with the highest energy in the three outcoming $\alpha$ particles, the right panel displays the distribution of the $\alpha$ particles with minimum energy, and the middle panel for the $\alpha$ with middle energy. Looking the context for details.

| Parameters | Values |
|---|---|
| Batch size | 32 |
| Learning rate | 0.0001 |
| Loss | SmoothL1Loss |
| epochs | 15 |
| Optimizer | Adam ($\beta_1 = 0.9, \beta_2 = 0.999$) |
| Weight decay | $\begin{cases} 0, & \text{for} \quad \text{VGG} - 16 \\ 10^{-3}, & \text{for} \quad \text{ResNet} - 18 \\ 10^{-4}, & \text{for} \quad \text{ResNet} - 34 \end{cases}$ |
| Scheduler type | ExponentialLR (optimizer, gamma=0.35) |

Table 6: Hyperparameters for models on energy reconstruction.

| Algorithm | $\sigma_{E_1}$ (keV) | $\sigma_{E_2}$ (keV) | $\sigma_{E_3}$ (keV) |
|---|---|---|---|
| VGG-16 | 114.9 | 136.2 | 104.6 |
| ResNet-18 | 68.2 | 76.1 | 76.2 |
| ResNet-34 | 69.3 | 73.6 | 67.0 |

Table 7: The standard deviation of energy reconstruction for models on the $^{12}$C test dataset.

To solve the issue of poor performance in angle reconstruction, we assumed that the $\alpha$ particle with the highest energy was produced by the $^{12}$C decay process, while the other two $\alpha$ particles were considered as decay products of $^8$Be.

Consequently, the momentum of $^8$Be was treated as the vector sum of the momenta of the two particles. Models were trained to extract relevant features from the dataset and reconstruct the emission angle of both $\alpha$ and $^8$Be. By applying preprocessing steps, we aimed to achieve higher accuracy in the emission angle reconstruction of these particles. As described in Section 3.2.1 and 3.2.2, the primary vertex and energy of three $\alpha$ particles can be reconstructed well by ResNet. Therefore, when the emission angle of $\alpha$ and $^8$Be can be predicted well, we can have the complete kinematic information for $^{12}$C decay events.

Fig. 9 presents the machine learning capabilities of ResNet-

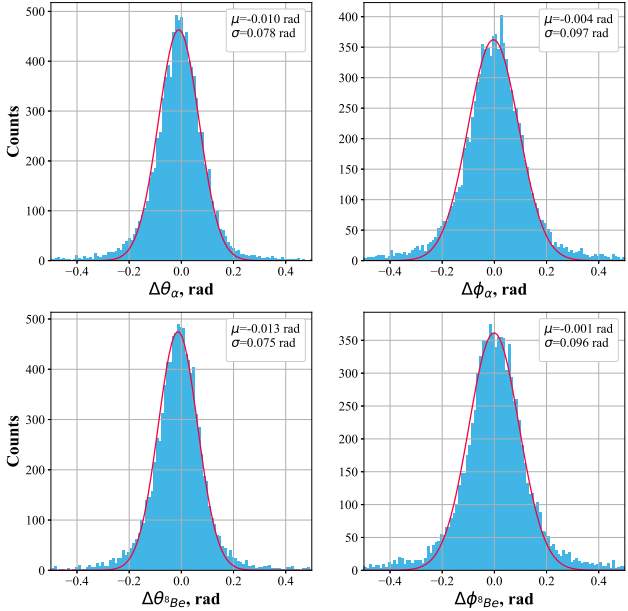

Fig. 9. (Color online) An example of angular reconstruction for outgoing particles. The upper panel shows the reconstruction result for $\alpha$ while the bottom panel is the angular difference(prediction-true) distribution of $^8$Be. The predictions are produced by ResNet-18.

| Parameters | Values |
|---|---|
| Batch size | 32 |
| Learning rate | 0.0001 |
| Loss | SmoothL1Loss |
| epochs | 20 |
| Optimizer | Adam ($\beta_1 = 0.9, \beta_2 = 0.999$) |
| Weight decay | $\begin{cases} 0, & \text{for } \text{VGG}-16 \\ 10^{-3}, & \text{for } \text{ResNet}-18 \\ 10^{-2}, & \text{for } \text{ResNet}-34 \end{cases}$ |
| Scheduler type | ExponentialLR (optimizer, gamma=0.35) |

Table 8: Hyperparameters for models on emission angle reconstruction.

| Particle | learning task | VGG-16 | ResNet-18 | ResNet-34 |
|---|---|---|---|---|
| $\alpha$ | $\theta$ (rad) | 0.198 | 0.078 | 0.092 |
|  | $\phi$ (rad) | 0.200 | 0.097 | 0.111 |
| $^8$Be | $\theta$ (rad) | 0.202 | 0.075 | 0.093 |
|  | $\phi$ (rad) | 0.196 | 0.096 | 0.108 |

Table 9: The standard deviation of angular reconstruction for models on the $^{12}$C test dataset.

18 in reconstructing the emission angles of $\alpha$ and $^8$Be particles. The upper panel of the figure displays the reconstruction results for $\alpha$ particles, while the bottom panel illustrates the reconstruction of $^8$Be particles. It is noteworthy that both $\alpha$ and $^8$Be particles' emission angles were simultaneously trained using CNN

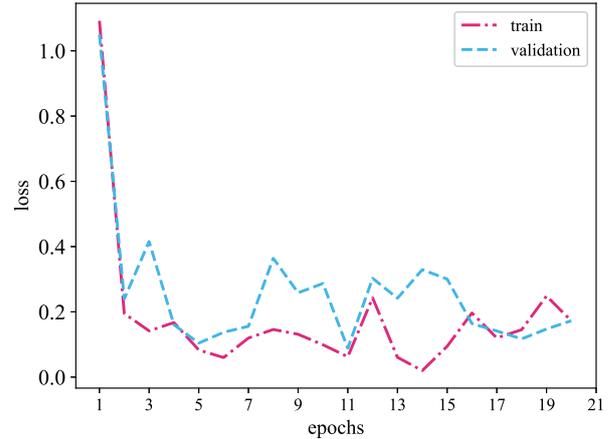

Fig. 10. (Color online)A representitive learning curve of ResNet-34, which indicates the absence of the inflection point of overfitting during the training process.

models.

The hyperparameters used during the training of models are listed in Table 8. A comparison of the three CNN models used for angular reconstruction is presented in Table 9, which demonstrates that ResNet outperforms VGG-16. For ResNet, the reconstruction deviation of $\theta$ and $\phi$ are both less than 0.1rad, indicating a high level of accuracy in the angular reconstruction.

As shown in Table 9, the training results of the two ResNet models are similar, but ResNet-18 performs slightly better than ResNet-34. This is a surprising result since ResNet-34, with a deeper network architecture, is expected to have better prediction capabilities. However, due to insufficient training, the learning curve did not reach the inflection point of overfitting, as shown in Fig. 10. To achieve the best performance of ResNet-34 in reconstructing the emission angles of outgoing particles, the hyperparameters used during training need to be adjusted several times until overfitting occurs. This training process is time-consuming. Therefore, in future research, a pre-trained ResNet-34 with fine-tuning can be used for angular reconstruction, improving the efficiency of the training process and enhancing the accuracy of particle angular reconstruction.

## 4. Conclusions

In this study, two Convolutional Neural Network architectures, VGG and ResNet models, are trained with data from Monte Carlo samples for event classification and reconstruction, where VGG-16, ResNet-18, and ResNet-34 were investigated. Our results demonstrate the effectiveness of machine learning methods in identifying $^{12}$C events from background, with ResNet-34 achieving an impressive precision of 0.99 on simulation data, which represents an almost perfect upper limit for the ResNet-34 classifier. It is of significant value for experimentalists that the actual data obtained from detectors can be automatically classified by CNN models. Furthermore, this approach can be easily adapted to other experiments that employ different instruments or detector technology, as long as an accurate physics simulator is accessible.

The effectiveness of CNN models in event reconstruction was also demonstrated through a comparison of the performance of VGG-16, ResNet-18, and ResNet-34. ResNet-18 was found to be the best performing CNN model, achieving almost similar accuracy with ResNet-34 in vertex, energy, and angular reconstruction, while consuming fewer computational resources. Specifically, ResNet-18 achieved an energy resolution of $\sigma_E < 77$ keV for alpha decay from $^{12}$C and an angular reconstruction deviation of $\sigma_\theta < 0.1$ rad.

In future research, our plan is to utilize the ResNet model trained on MC samples to classify and predict experimental data. However, as highlighted in Ref. [23], models trained on simulation data to classify experimental data may not perform as well as on simulation test datasets due to the lack of true structural noise in simulated data. Therefore, it is essential to extract the structural noise observed in experimental data and use it to generate simulated data with noise to enhance the performance of the model. Additionally, a pre-trained ResNet on ImageNet can be a feasible option for this research, saving significant time for model training and requiring only a small sample size for fine-tuning to achieve optimal performance.

## Acknowledgments

This work is supported in part by National Key R&D Program of China No. 2022YFA1602402, 2020YFE0202001 and the National Natural Science Foundation of China (NSFC) under Grant No. 12235003, 11835002, 11705031, and 12147101.